\documentclass[11pt,preprint2]{aastex}
\received{17 Dec. 2002}
\accepted{4 Aug. 2003}
\slugcomment{To appear in Astrophysical Journal}
\shorttitle{Infrared Spectroscopy of Radio Galaxies}
\shortauthors{Iwamuro et al.}
\begin{document}
\title{INFRARED SPECTROSCOPY OF 15 RADIO GALAXIES AT $2<z<2.6$}
\author{\sc Fumihide Iwamuro\altaffilmark{1}, Kentaro Motohara\altaffilmark{2}, Toshinori Maihara\altaffilmark{1}, Masahiko Kimura\altaffilmark{1}, Shigeru Eto\altaffilmark{1}, Takanori Shima\altaffilmark{1}, Daisaku Mochida\altaffilmark{1}, Shinpei Wada\altaffilmark{1}, Satoko Imai\altaffilmark{1}, and Kentaro Aoki\altaffilmark{3}}
\altaffiltext{1}{Department of Astronomy, Kyoto University, Kitashirakawa, Kyoto 606-8502, Japan.}
\altaffiltext{2}{Institute of Astronomy, School of Science, University of Tokyo, 2-21-1 Osawa, Mitaka, Tokyo 181-0015, Japan.}
\altaffiltext{3}{Subaru Telescope, National Astronomical Observatory of Japan, 650 North A`ohoku Place, University Park, Hilo, HI 96720.}
\begin{abstract}
Near-infrared spectra of 15 high-redshift radio galaxies (HzRGs) located 
at $2<z<2.6$ were obtained by the OH Airglow Suppressor spectrograph
mounted on the Subaru telescope. The UV-optical line ratio diagnostic diagrams
indicate that half of the observed HzRGs have extended emission-line 
regions with low metal abundance, photoionized by a flat-continuum active galactic nucleus such as a quasar. 
We also found two probable correlations between radio and rest-optical parameters: 
(1) HzRGs with massive hosts tend to have a redder rest-optical continuum, and (2) HzRGs 
with smaller radio sizes also show a redder optical continuum. On the basis of the 
correlations, the nature of HzRGs at $2<z<2.6$ is discussed.
\end{abstract}

\section{INTRODUCTION}
High redshift radio galaxies (HzRGs) have been studied as progenitors of 
the present massive galaxies. Their rest-UV morphologies usually show 
a clumpy structure with a bright nucleus \citep{rus97,pen99}, aligned 
with the axis of the radio sources in the so-called alignment effect. The major 
cause of the alignment effect is considered to be one of the following three
mechanisms: (1) induced star formation by the passage of the radio jet \citep{
mcc87, cha87}, (2) scattered light from a hidden active galactic nucleus (AGN: 
di Serego Alighieri et al. 1989; Cimatti et al. 1993), and 
(3) the nebular continuum \citep{dic95}. These stellar, AGN, and nebular 
components make a nearly equal contribution to the UV continuum at the 
nucleus region of intermediate-redshift radio galaxies \citep{tad02}.
The rest-optical imaging of HzRGs shows the morphological evolution 
of host galaxies: multiple components spread over $\sim$100~kpc at $z>3$, 
a single compact structure dominating at $2<z<3$, and an elliptical profile
with a de Vaucouleurs r$^{1/4}$ law formed at $z\sim 1$ \citep{van98, pen01}.
The morphologies of the extended emission-line regions (EELRs) indicate the
existence of a photoionization cone or shock-ionized region extended to 
several tens of kiloparsecs \citep{arm98,mot00,ega03}. They strongly support the 
unification model between radio loud quasars and radio galaxies.

The characteristics of HzRGs have been investigated by rest-UV spectroscopy.
An HzRG with a smaller radio size has dense \ion{H}{1} clouds surrounding
the inner Ly$\alpha$ emitting region with a large turbulent velocity, 
probably because of the environmental effect on the radio size \citep{van97}.
The total (nucleus+EELR) ionization mechanism of UV emission lines is well 
explained by photoionization with a power-law index of $\alpha=-1.0$ 
\citep{vil97}, except for \ion{C}{2}], which is enhanced by shock \citep{deb00}, and 
\ion{N}{5}, which indicates an overabundance of nitrogen (Villar-Mart{\'i}n et al. 1999, 2001; Vernet et al. 2001)
Concerning the physical conditions of the EELR, the existence of a metallicity gradient 
\citep{ove01} and the contribution of shock in the outer halo \citep{tan01} were 
reported, but the general mechanisms are still unknown. The contribution of
the scattered AGN light to the UV continuum varies from object to object, 
between $\sim 15\%$ and $\sim 85\%$, estimated by spectropolarimetry 
\citep{ver01}.

On the other hand, rest-optical spectra of HzRGs show strong [\ion{O}{3}]
emission lines whose intensity has a correlation with the radio luminosity 
\citep{eal93,iwa96,eva98}. Reddened quasar light (or broad H$\alpha$) has 
also been detected from a few HzRGs \citep{lar00, fos00}. The emission line 
ratios of individual knots around the nucleus indicate that shock 
or starburst is the major ionization mechanism of these off-nuclear knots 
\citep{car01}. Although the rest-optical spectra of HzRGs became observable
within a reasonable exposure time by the newly built infrared instruments 
on the large telescopes, the published data are not sufficient to make 
statistical studies. 

In this paper we present the rest-optical spectra of 15 HzRGs at $2<z<2.6$ 
obtained by the OH Airglow Suppressor (OHS) spectrograph \citep{iwa01} mounted on 
the Subaru telescope. The typical wavelength coverage of 3400--5400~\AA\  
includes major emission lines of AGNs, such as [\ion{O}{2}]~$\lambda$3727, 
\ion{He}{2}~$\lambda$4686, H$\beta$, and [\ion{O}{3}]~$\lambda$5007. 
In $\S2$ we describe the observations and the reduction procedure 
of the 15 HzRGs. In $\S3$, after we summarize the observed emission-line properties, 
we examine the major ionization mechanisms using UV-optical line ratio diagnostic 
diagrams. In $\S4$ we discuss the nature of HzRGs on the basis of the probable 
correlations between several observed parameters, and we present our conclusions 
in $\S5$.

Throughout the paper we adopt a cosmology with $H_0=65$~km~s$^{-1}$~Mpc$^{-1}$, 
$\Omega_M=0.3$, and $\Omega_{\Lambda}=0.7$.

\section{OBSERVATIONS AND DATA REDUCTION}
A total of 15 well-known HzRGs with redshifts at $2<z<2.6$ were observed 
during five test observation runs of the OHS mounted on the Subaru telescope in 
2000 May, July, September, and December and 2001 August. For many targets, the 
rest-UV (and a part of the rest-optical) spectra and {\it Hubble Space Telescope} 
({\it HST}) and Very Large Array images are 
available from the past literature. The rest-optical spectroscopy was carried
out with the slit along the radio axis. The typical exposure sequence is 
1000~s in four positions, in which the object is moved about 10\arcsec\ 
along the slit by nodding the telescope after every exposure. The limiting 
magnitude is $H$=21.1 (S/N=5 for 4000~s) at best with this sequence. A nearby 
Smithsonian Astrophysical Observatory (SAO) star with a spectral type of A or F was observed just after this exposure 
sequence and was used as a spectroscopic standard to remove telluric atmosphere 
absorption features and to correct the instrumental response. The $J$- 
(1.108--1.353~$\mu$m) and $H$-band (1.477--1.804~$\mu$m) spectra were 
simultaneously obtained with a spectral resolution of $\sim$210. A slit width 
of 1\arcsec\ corresponds to 9 pixels on a pixel scale of 0\farcs111 in the 
case of the optical secondary mirror of the telescope, which was generally 
used during these observations. Since the typical seeing size of 0\farcs7 
is smaller than the slit width, we must take the morphology of the object 
within the slit into account when we consider the velocity structure of 
the emission lines. The observation log is summarized in Table~1.

All data reduction was performed using IRAF as follows:
\begin{enumerate}
\item Sky subtraction for a 
pair of spectroscopic images taken at different positions.
\item Flat-fielding using 
a standard $H$-band flat image (for correction of the local variations in 
the quantum efficiency).
\item Correction of unexpected hot pixels and cosmic-ray events.
\item Straightening of residual-airglow emission lines along column pixels.
\item Residual-airglow subtraction by fitting to each column excluding object 
positions (plus and minus images).
\item Shift-and-add of all images.
\item Correction of 
atmospheric absorption features and instrumental responses using the spectrum 
of the SAO star taken just after pointing at the object (where correction has already been 
applied to the hydrogenous absorption features and the intrinsic slope in the 
SAO spectrum).
\item Straightening of the object spectrum along line pixels.
\item Extraction of the object spectrum using a one-dimensional Gaussian mask with an
FWHM of 3\arcsec, except for TXS~0211$-$122, B3~0731$+$438, 4C~+10.48, 4C~+10.48 
(FWHM of 4\arcsec), and 4C~+23.56 (FWHM of 2\arcsec).
\end{enumerate}
The resulting one-dimensional 
spectrum was calibrated using the photometric results of the $H$-band slitless 
imaging frame taken just before the spectroscopic observation. As a result, the 
average flux of the $H$-band spectrum corresponds to the $H$-band magnitude 
within an aperture diameter of 2\farcs2 (20 pixels).

\section{RESULTS}
\subsection{Line Fluxes and Continuum Levels}
The final flux-calibrated spectra are presented in Figure~1. The right panels 
of this figure show the continuum profiles with 30~pixel smoothing, after 
fitting and subtraction of the emission lines with Gaussian functions.
In this fitting procedure, each emission line is restored, fitted again, 
and subtracted individually to improve the estimation of the local continuum 
level (which is often affected by the residuals of neighboring emission lines).
In order to achieve higher fitting accuracy, we assume a constant line ratio 
as well as the same FWHM and redshift for the following neon doublets: 
3.32 for [\ion{Ne}{3}]~$\lambda\lambda$3869,~3967 and 
2.74 for [\ion{Ne}{5}]~$\lambda\lambda$3426,~3346, which are calculated by
CLOUDY94 \citep{fer00}. Since both doublets are emitted by the transition 
from the $^1D_2$ to $^3P_{2,1}$ level, these ratios are expected to be invariable.
For faint emission lines, we assume that the redshift is the same as 
that for [\ion{O}{3}]~$\lambda$5007 and that the FWHM is the instrumental value 
of $\sim$75~\AA. The emission line properties and the continuum levels are 
listed in Table~2.

\subsection{Emission Line Ratios}
\subsubsection{UV-optical diagnostic diagrams}
The UV line fluxes of \ion{C}{4}~$\lambda$1549, \ion{He}{2}~$\lambda$1640, and 
\ion{C}{3}]~$\lambda$1909 are available for 13 out of 15 HzRGs from the past 
literature (Table~3). We combine these data with the optical line flux in 
Table~2 to examine the major ionization mechanisms of HzRGs. Figure~2 shows 
the UV-optical line ratio diagnostic diagrams. We calculated $4\times 4\times 4$ 
photoionization sequences with CLOUDY94 \citep{fer00} with power-law spectral 
indexes $\alpha=0.0,-0.5,-1.0,$ and $-1.5$, metal abundances $Z=0.1,0.3,1.0,$ and 3.0~$Z_{\odot}$, 
and hydrogen densities $N_{\rm H}=10,100,1000,$ and $10^4~$cm$^{-3}$. The case of 
$N_{\rm H}=1000~$cm$^{-3}$ is plotted in Figure~2. Since the contribution 
of the hydrogen density to the emission line ratio is smaller than that of the 
other parameters, we can estimate the rough trend of $\alpha$ and $Z$ 
only from this case. The grid of the shock(+precursor) models \citep{dop95,all99,deb00} 
is also plotted in Figure~2.

The [\ion{O}{2}]/[\ion{O}{3}] vs. [\ion{O}{3}]/H$\beta$ diagram in Figure~2 is one of the most
useful diagnostic diagrams for estimating the ionization parameter log$U$, while 
the [\ion{O}{2}]/[\ion{O}{3}] vs. \ion{C}{4}/\ion{C}{3}] and \ion{C}{4}/\ion{He}{2} vs. 
[\ion{O}{3}]/H$\beta$ diagrams have advantages for estimating $\alpha$ and $Z$,
respectively. The degeneracy of the parameters can be solved by these diagrams,
indicating $\alpha\sim -0.5$, $Z\sim 0.2\ Z_{\odot}$, and log$U$ $\sim -2.2$
for our sample. On the other hand, the shock+precursor models are completely separated 
from the photoionization models in the [\ion{O}{2}]/[\ion{O}{3}] vs. \ion{He}{2}/H$\beta$
diagram. The majority of our samples seem to agree with the photoionization models; 
however, we cannot rule out the shock+precursor models for several objects only 
by this figure. The dust extinction, which may have a significant effect on the 
scatter of the data in the top left diagram of Figure~2, is expected to be small, 
because (1) the sizes of the EELRs are very large (10--30~kpc), and (2) the color of the 
rest-UV continuum \citep{rot97,ver01} is blue.

\subsubsection{Best fit parameters}
To determine the best-fit parameters for each object, we considered a 
nine-dimensional line ratio diagnostic diagram, which consists of all the UV/UV combinations of 
three UV lines (\ion{C}{4}, \ion{He}{2}, and \ion{C}{3}]: three combinations) and all the 
optical/optical combinations of four optical lines ([\ion{O}{2}], H$\beta$, \ion{He}{2}, 
and [\ion{O}{3}]: six 6 combinations). Here we did not use any UV-optical line ratios, 
which may be affected by the differences of the instruments and the other observation 
parameters. In this nine-dimensional diagnostic diagram, the model grids as shown in Figure~2 were 
divided into 10 subgrids at even intervals. After the most probable subgrid giving 
the minimum $\chi^2$ value was found, the nearest subgrids were divided into 10 
sub-subgrids in order to search the most probable sub-subgrids again. The best-fit 
model with parameters was determined by iteration of the above calculation, as listed in 
Table~4. More than half of the objects can be explained by a photoionization model with a 
flat ionizing continuum and low metallicity, consistent with the distribution of the data 
in Figure~2. On the other hand, four objects are (also) in good agreement with the 
shock+precursor model, while three objects cannot be explained by a single-photoionization 
or shock+precursor model with a significant confidence level. 

\section{DISCUSSION}
\subsection{Correlations between Parameters}
Before we discuss the nature of HzRGs, we review the correlations between the radio 
parameters (size, power, and spectral index) and the observed rest-optical parameters (continuum
brightness, spectral index, [\ion{O}{3}] luminosity, and FWHM). The significance levels at 
which the null hypothesis of zero correlation is disproved are listed in Table~5 together 
with the correlation coefficients. Values marked with an asterisk in Table~5 indicate the probable correlations.
In $\S4.2$ we discuss these correlations individually.

\subsection{Probable Correlations}
\subsubsection{Correlation between $P_{opt}$ and $\alpha_{opt}$}
The correlation of the rest-optical continuum slope between [\ion{O}{2}]~$\lambda$3727 and 
[\ion{O}{3}]~$\lambda$5007 ($\alpha_{opt}$; see Figure~1) with the continuum brightness 
at the rest-5007~\AA\ wavelength ($P_{opt}$) is shown in the left panel of Figure~3. 
Here $P_{opt}$ roughly represents the mass of the host galaxy, because most of the 
continuum flux at the rest-5007~\AA\ wavelength is expected to be of stellar origin (e.g., Fosbury 2000). 
With this taken into account, the $P_{opt}$-$\alpha_{opt}$ correlation indicates that 
a HzRG with a massive host has a redder optical color. For reference, the $\alpha_{opt}$ 
values of the instantaneous starburst model change from $-$0.2 (10~Myr) via $-$1.7 (100~Myr) 
to $-$4.1 (1~Gyr) without dust extinction, while the value of the scattered AGN light is 
expected to be $\alpha_{opt}\sim$0.

\subsubsection{Correlation between $D_{radio}$ and $\alpha_{opt}$}
The correlations of $\alpha_{opt}$ with the radio size ($D_{radio}$) are shown 
in the right panel of Figure~3. A radio source with a smaller $D_{radio}$ is 
usually thought to have a younger age of radio activity \citep{blu99} or/and have dense gas 
clouds that interrupt the expansion of the radio jet (O'Dea 1998 and references therein). 
On the basis of the former idea, the older radio source should inhabit the bluer host galaxy, 
which seems difficult to explain. The latter idea indicates that the redder HzRG not only 
has a massive host (see $\S4.2.1$) but also tends to be surrounded by dense gas clouds. 
Although the significance level of the $D_{radio}$-$P_{opt}$ correlation does not reach 
a high confidence level (see Table~5), this inverse correlation does not conflict with this idea.
The age of the radio source probably contributes to the scatter in these correlations.

\subsection{Nature of HzRGs at $2<z<2.6$}
From the UV-optical diagnostic diagrams and the probable correlations discussed above, 
the nature of HzRGs located at $2<z<2.6$ is summarized as follows: 
The HzRGs with massive hosts show a redder optical color, indicating that the scattered light 
of the AGNs or massive blue stellar population does not make a significant contribution to their
rest-optical spectra. These galaxies are also surrounded by dense gas clouds that interrupt the 
expansion of the radio jet. The radio activities do not have a large effect on these massive hosts, 
because the ages of the host galaxies expected from their red optical color ($\gtrsim$100~Myr) 
are older than the ages of typical radio activities ($\sim$10~Myr). On the other hand, the HzRGs with 
less massive hosts are not surrounded by dense gas clouds, corresponding to a larger radio 
size. The considerable contribution of the scattered light of the AGNs or massive blue stellar 
population makes their optical color blue. From the results of rest-UV spectropolarimetry 
reported by \citet{ver01}, the scattered light is the dominant contribution to these blue HzRGs
with larger radio sizes (while massive red HzRGs such as B3~0731$+$438 and 4C~+40.36 have a 
smaller contribution from scattered light). The ages of the host galaxies expected from their 
blue optical color ($\lesssim$100~Myr) are almost comparable to the ages of radio activities, 
implying that the current radio activities are closely related to the formation of the 
host galaxies.

The EELRs of HzRGs are usually photoionized by a flat-continuum AGN such as a quasar ($\alpha\sim -0.5$),
and the metal abundance is expected to be subsolar ($Z\sim 0.2\ Z_{\odot}$). In a few objects, 
shock+precursor is the dominant ionization mechanism of the EELR, and three out of 13 objects 
cannot be explained by the single-ionization mechanism. A likely explanation is that the 
ionization mechanism of the EELR varies regionally, which requires further investigation 
into the spatial variation of the emission line ratios as well as the combination of 
photoionization and shock+precursor models \citep{deb00, moy02} to fit the entire EELR 
spectra.

\section{CONCLUSIONS}
The nature of HzRGs located at $2<z<2.6$ was investigated using their rest-optical spectra 
obtained by the OHS mounted on the Subaru telescope. We found two probable correlations 
between the radio and rest-optical parameters of the observed HzRGs as well as the characteristics 
of their ionization condition using UV-optical diagnostic diagrams.
\begin{enumerate}
\item {\it Correlation of }$P_{opt}$-$\alpha_{opt}$.--- HzRGs with massive hosts tend to have a redder optical continuum, 
indicating that the contribution of the scattered light of the AGNs or massive blue stellar 
population to the rest-optical spectra of these massive HzRGs is quite small. The contribution 
of the scattered light becomes larger for less massive objects, making their color blue.
\item {\it Correlation of }$D_{radio}$-$\alpha_{opt}$.--- HzRGs with smaller radio sizes also show a redder optical continuum.
Together with the above correlation, HzRGs with massive hosts are surrounded by dense gas clouds, 
which interrupt the expansion of the radio jets. The radio activities do not have a large effect on 
these massive hosts, while they are closely related to the formation of the smaller host galaxies.
\item UV-optical diagnostic diagrams.--- More than half of the examined HzRGs are photoionized by a 
flat-continuum AGN such as a quasar ($\alpha\sim -0.5$), which supports the unification model between 
HzRGs and quasars. In this case, the metal abundance of the EELR is subsolar ($Z\sim 0.2\ Z_{\odot}$).
The shock+precursor models can explain a few cases, being expected to contribute to three other
cases, which cannot be fitted by a single model.
\end{enumerate}

\acknowledgments
These results were accomplished during test observation runs of the OHS using the Subaru 
telescope. We are therefore indebted to all members of the Subaru Observatory, National Astronomical 
Observatory of Japan. 

\appendix
\section{Notes on Particular Objects}
Although most objects have a compact nucleus surrounded by an EELR, some of 
them show knotlike components or significant velocity structure. In this 
appendix, we show the two-dimensional spectra of these particular objects 
(Figures~4 and 5) and discuss their features individually.

\subsection{MRC~0156$-$252}
This spectrum consists of a compact red continuum and [\ion{O}{3}] 
emission lines. Since the position angle of the slit (50$^{\circ}$) is almost 
the same as the position angle of the $H$-band morphology of this galaxy taken by 
{\it HST}/NICMOS \citep{pen01}, the origin of the tilt of the [\ion{O}{3}] 
emission lines is not a morphological effect within the slit but a 
redshifted velocity structure of $\sim$500~km~s$^{-1}$. This [\ion{O}{3}]
jet extends $1\arcsec$ southwest along the slit from the nucleus. Our result is 
consistent with the dusty-quasar nature of this object reported by 
\citet{eal96}.

\subsection{MRC~0200+015}
This galaxy has an [\ion{O}{3}] knot separated 2\farcs 2 south-southeast from the 
nucleus, which probably corresponds to the south radio lobe of this object
\citep{car97}.

\subsection{MRC~0214+183}
The detailed morphology of this galaxy is unknown. If the inclination of 
emission lines is due to the velocity difference, this galaxy has bipolar
jets with velocities of approximately +400 and $-$600~km~s$^{-1}$ to the north and south 
directions, respectively.

\subsection{MRC~0406$-$244}
This spectrum shows multiple components of emission lines with a single 
continuum. The main body of this galaxy consists of two components in 
the optical and near-infrared images taken by {\it HST} \citep{rus97,pen01}, 
which correspond to components a$_1$ and a$_2$ in Figure~5. Component a$_1$ is 
located at the ridge of the continuum emission of the host galaxy.
Component a$_2$ is a redshifted knot with a velocity of $\sim$500~km~s$^{-1}$.
The other two knots (components b and c in Figure~5) are parts of the outer EELR 
of this galaxy.

\subsection{B3~0731+438}
The knotlike structure of the [\ion{O}{3}] emission lines are a part of the 
ionization cone of this galaxy \citep{mot00}. The velocity offset of this 
component is not real but due to the morphological structure within the slit.
Most of the line emission comes from the southern ionization cone, where 
a faint blue continuum is marginally detected, indicating the existence of a 
blue stellar population rather than scattered AGN light \citep{ver01}.

\subsection{TXS~0828+193}
This galaxy has the most complicated spectrum in our sample. The [\ion{O}{3}]
emission lines consist of three components, which correspond to the knots 
b, a, and c in Figure~5 of \citet{ste02} from top to bottom. We cannot confirm a velocity 
structure of [\ion{O}{3}] emission lines, because it is probably caused by 
a morphological effect within the slit. The continuum emission shows two 
components: one is a bluer northern component including [\ion{O}{3}] knot b, 
and the other is a redder southern component including [\ion{O}{3}] knot a. 
These results strongly support the idea that the northern component is
the scattered light from a hidden AGN in the southern host galaxy 
(Pentericci et al. 1999; Steinbring et al. 2002).

\subsection{4C~$-$00.62}
The EELR of this galaxy extends along the radio jet to the west-southwest direction 
\citep{pen00}. 

\subsection{4C~+40.36}
This spectrum shows the most broad [\ion{O}{3}] emission-line width in our sample, 
which is more than 1400~km~s$^{-1}$ after the instrumental contribution is 
subtracted. The distortion of emission lines is probably due to the morphological 
structure of this object \citep{ega03}.

\clearpage
\onecolumn
\begin{figure}
\plotone{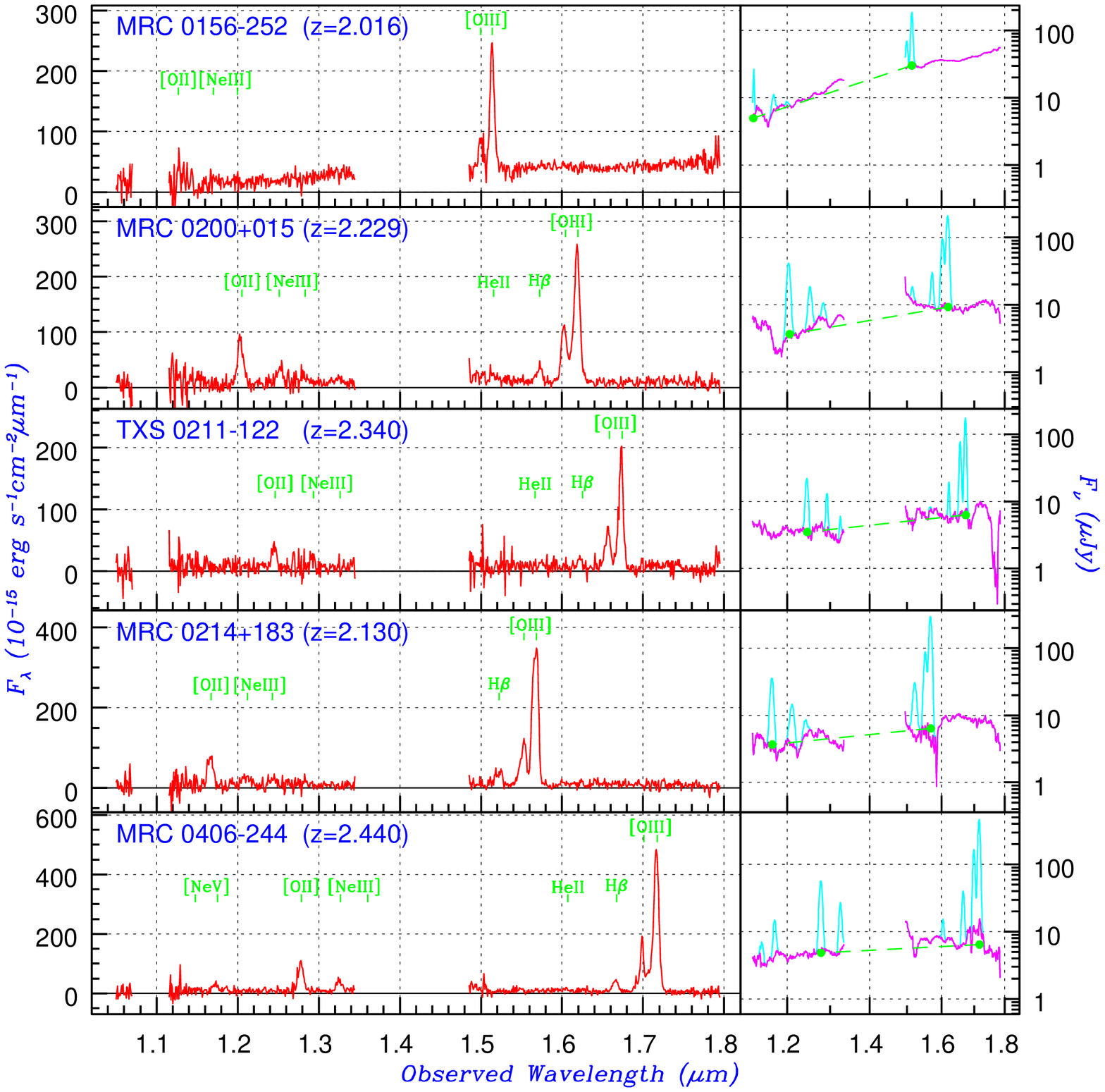}
\caption{One-dimensional spectra of $2<z<2.6$ HzRGs. The flux is calibrated 
using the $H$-band magnitude with a 2\farcs2 diameter aperture (see section 2). 
The right panels show the continuum shape smoothed over 30 pixels ({\it thick lines}) 
after subtraction of the emission lines ({\it thin lines}). The continuum levels at 
[\ion{O}{2}]~$\lambda$3727 and [\ion{O}{3}]~$\lambda$5007 (listed in Table~2) 
are indicated by filled circles and connected with dashed lines. [{\it See the 
electronic edition of the Journal for a color version of this figure.}]}
\end{figure}
\setcounter{figure}{0}
\begin{figure}
\plotone{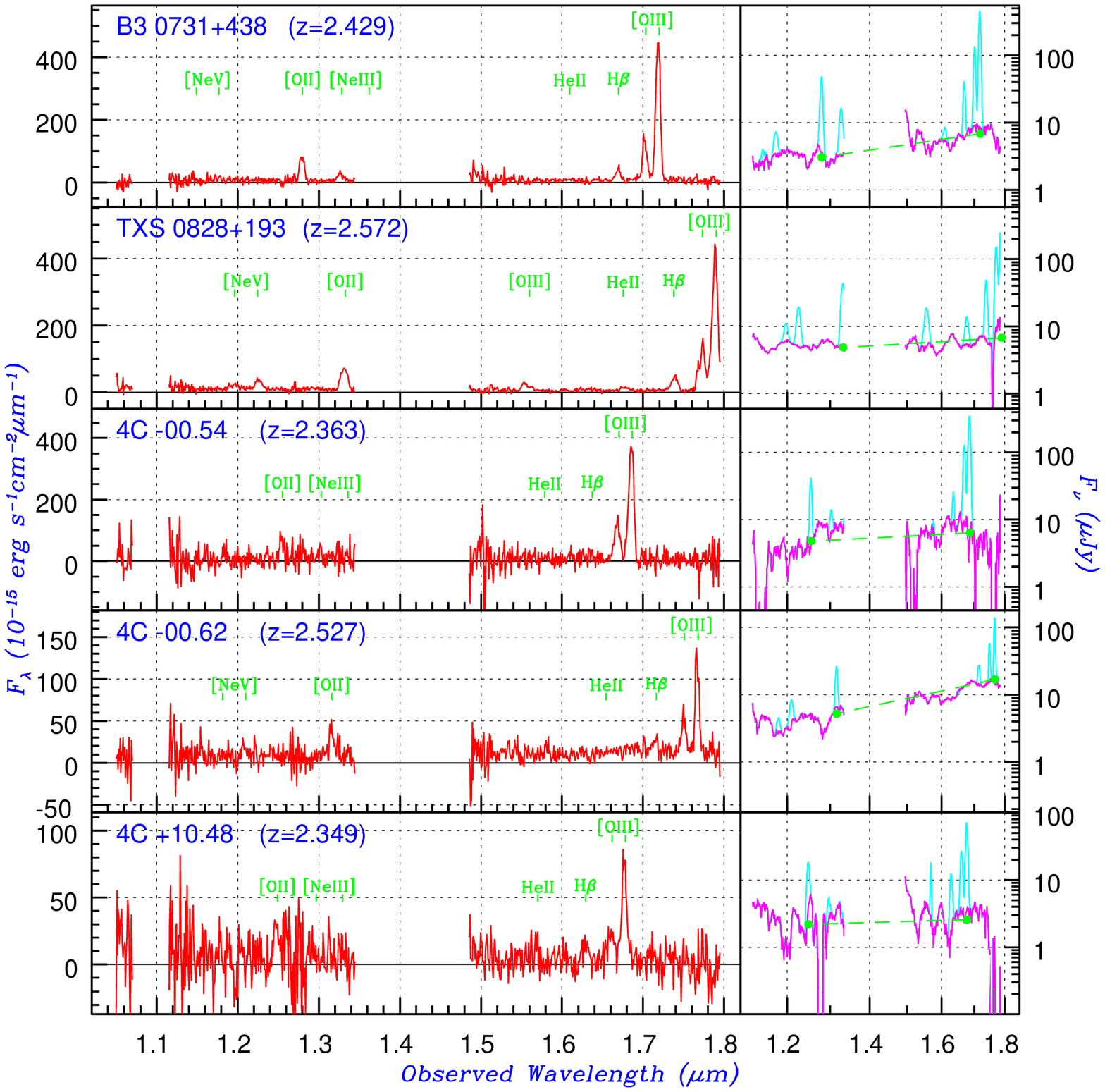}
\caption{(continued)}
\end{figure}
\setcounter{figure}{0}
\begin{figure}
\plotone{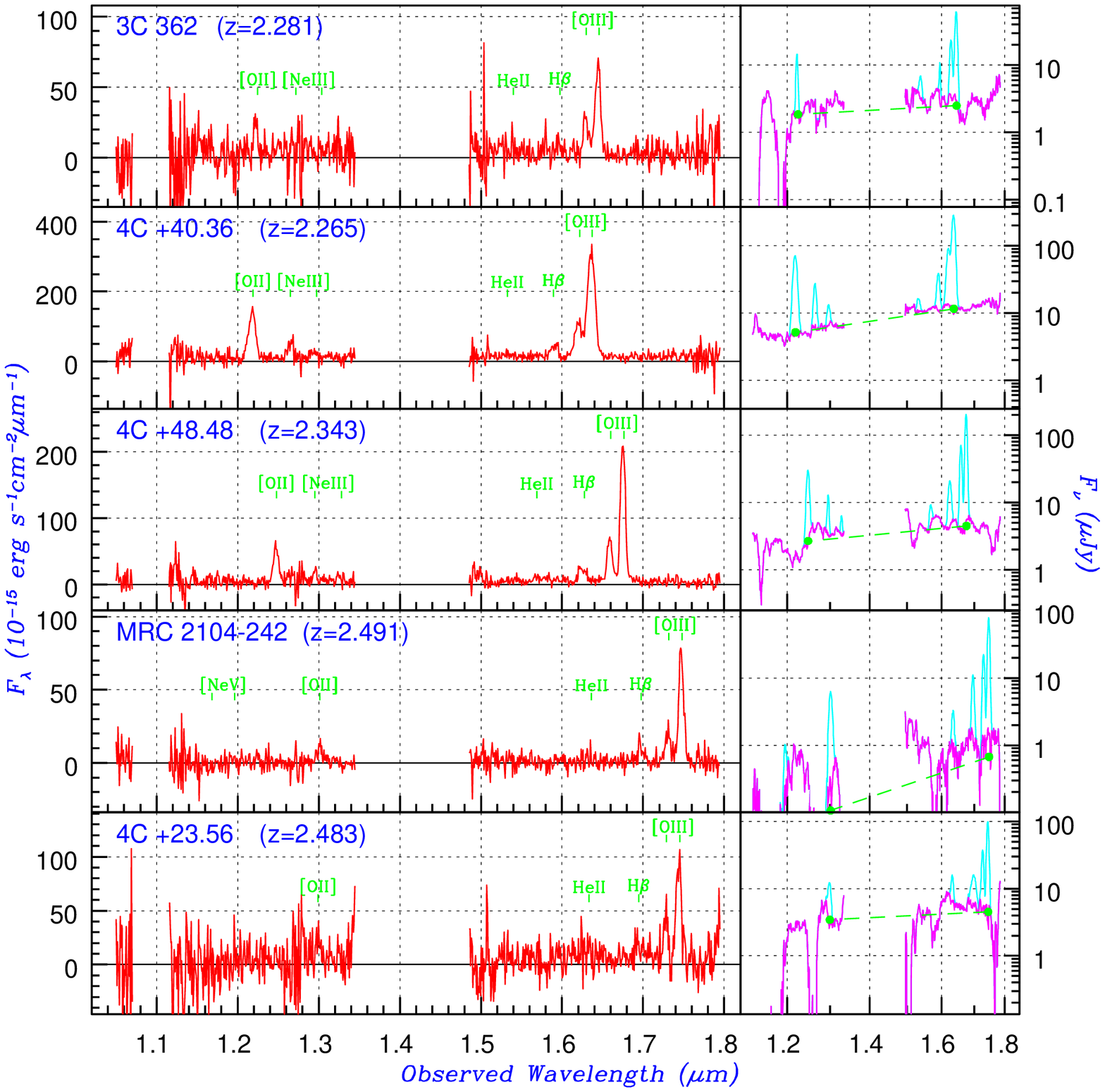}
\caption{(continued)}
\end{figure}
\begin{figure}
\plotone{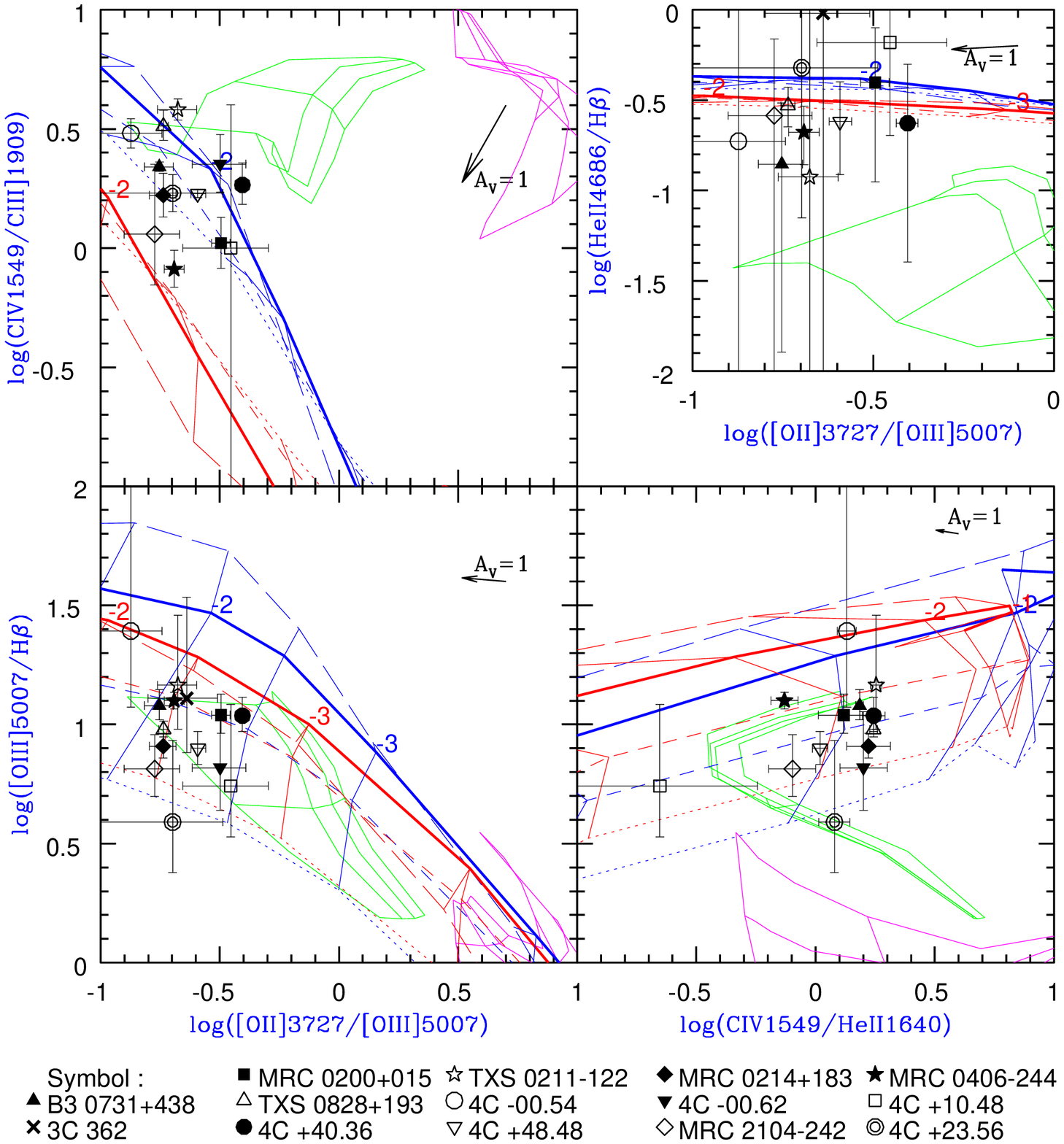}
\caption{Optical-UV line ratio diagnostic diagrams. Plots of $2\times 4$ photoionization 
sequences with power-law spectral indexes of $-0.5$ ({\it blue lines}) and $-1.0$ ({\it red lines}) and with metal 
abundance $Z=0.1$ ({\it dotted lines}), 0.3 ({\it dashed lines}), 1.0 ({\it thick solid lines}), and 
3.0~$Z_{\odot}$ ({\it long-dashed lines}) are plotted with a 0.5 dex step of log$U$. 
The hydrogen density for these models is the fixed value of $N_{\rm H}=1000$~cm$^{-3}$. Shock ({\it magenta lines}) 
and shock+precursor ({\it green lines}) models are shown with a grid of $B/\sqrt{n}=0, 1, 2,$ and 4~$\mu$G~cm$^{3/2}$ 
and $v=200, 300,$ and 500~km~s$^{-1}$.}
\end{figure}
\begin{figure}
\plotone{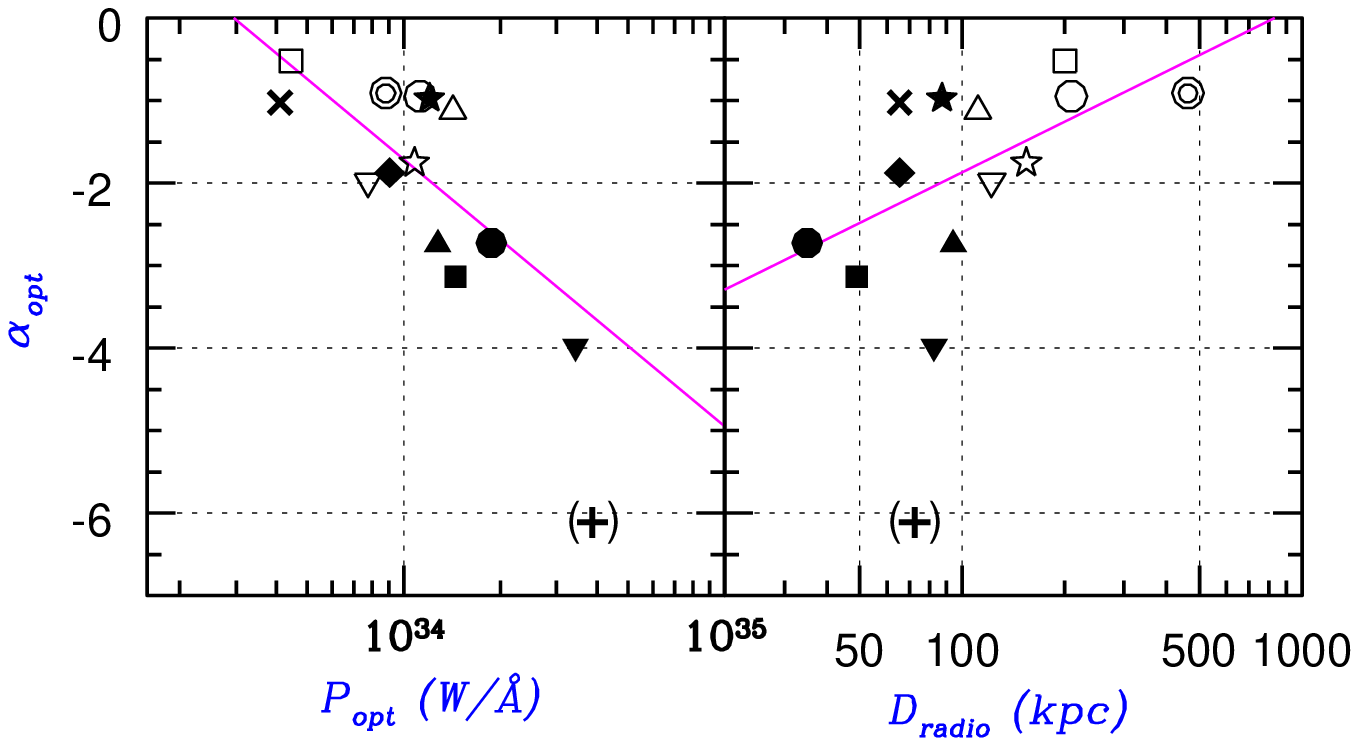}
\caption{Probable correlations between parameters. The symbols are the same as in Figure~2, while solid lines correspond to the results of linear least-squares fits. MRC~0156$-$252 ({\it plus symbol with brackets}) is plotted together with the other samples only for comparison and is not included in the estimation of the correlation significance. MRC~2104$-$242 is also excluded from the estimation of the correlation significance with $\alpha_{opt}$ (see footnotes of Table~5). [{\it See the electronic edition of the Journal for a color version of this figure.}]}
\end{figure}
\begin{figure}
\caption{Two-dimensional spectra of particular objects. The observed spectra have discontinuous spectral ranges because of the instrumental throughput of OHS: 1.05--1.08, 1.11--1.35, and 1.48--1.80~$\mu$m from left to right. [{\it See the electronic edition of the Journal for a color version of this figure.}]}
\end{figure}
\begin{figure}
\plotone{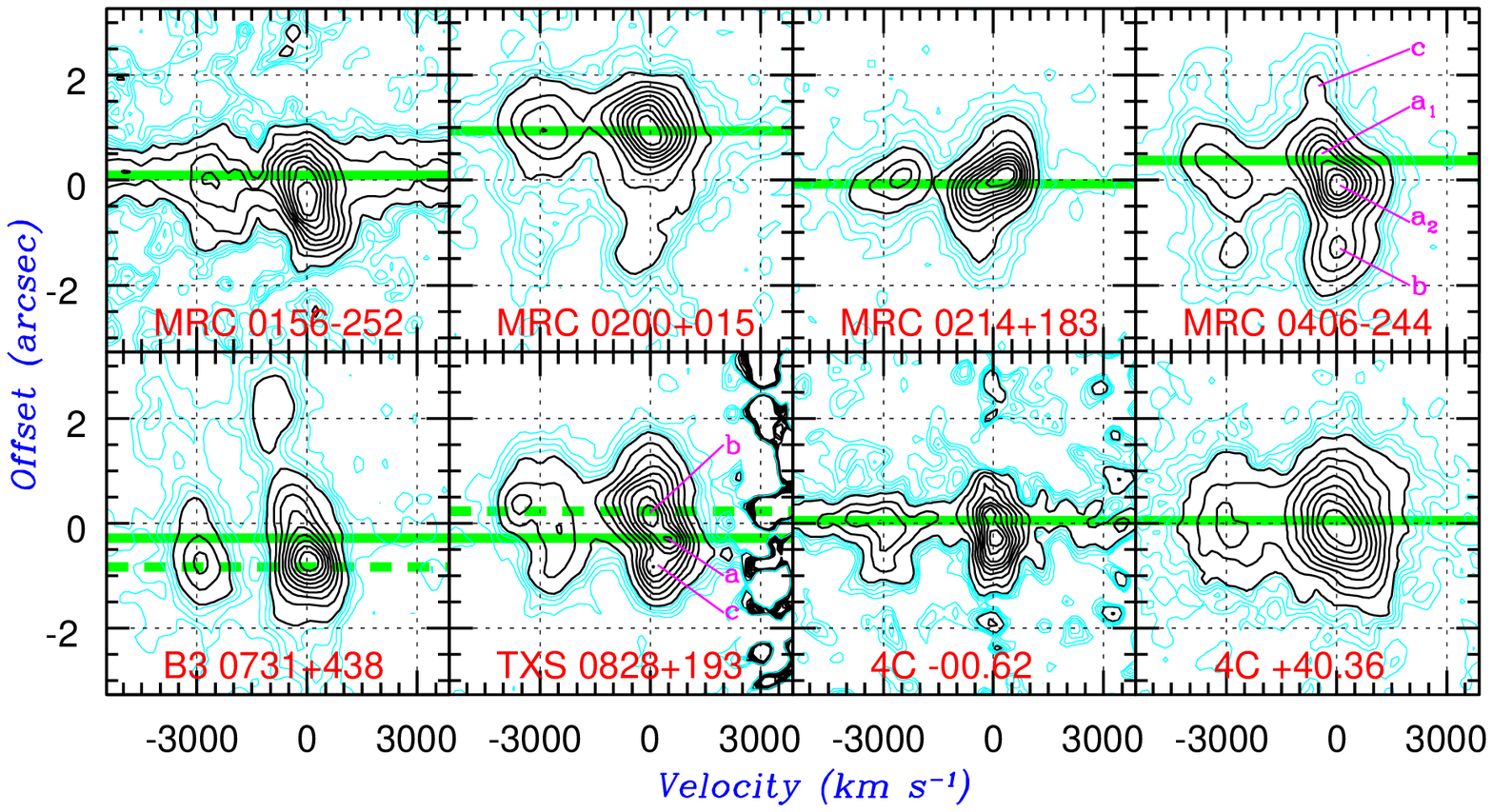}
\caption{Two-dimensional emission line structures of the [\ion{O}{3}]~$\lambda
\lambda$4959,5007 shown in Figure~4. The spectra have been smoothed with a 
Gaussian of 1 pixel (0\farcs111 or 8.7~\AA). Thick contours are linearly 
spaced from 10\% to 90\% of the peak flux, with an interval of 10\%, while the 
thin contours denote 2\%, 4\%, 6\%, and 8\% levels. Solid horizontal lines show the 
position of the continuum emission of the host galaxies. Dashed horizontal 
lines in B3~0731+438 and TXS~0828+193 correspond to the position of the 
faint blue continuum (see Appendix). [{\it See the electronic edition of the Journal for a color version of this figure.}]}
\end{figure}
\clearpage
\begin{deluxetable}{lccccccc}
\tabletypesize{\normalsize}
\tablecaption{\sc Log of Observations}
\tablewidth{0pt}
\tablehead{\colhead{} & \colhead{} & \colhead{} & \colhead{} & \colhead{} & \colhead{Exposure} & \colhead{Slit P.A.} & \colhead{Seeing}\\
\colhead{Common Name} & \colhead{IAU Name} & \colhead{$H$-mag\tablenotemark{a}} & \colhead{Redshift} & \colhead{Date} & \colhead{Time (s)} & \colhead{(degree)}& \colhead{(arcsec)}}
\startdata
\objectname[]{MRC~0156$-$252} &0156$-$252& 18.45 & 2.016 & 2000 Sep 15 & 4000 & $+$50 & 0.6 \\
\objectname[]{MRC~0200$+$015} &0200$+$015& 19.26 & 2.229 & 2001 Jul 31 & 3000 & $-$24 & 0.6 \\
\objectname[]{TXS~0211$-$122} &0211$-$122& 19.82 & 2.340 & 2000 Dec 18 & 4000 & $-$78 & 0.7 \\
\objectname[]{MRC~0214$+$183} &0214$+$183& 19.19 & 2.130 & 2001 Aug 01 & 5000 &  $-$2 & 0.6 \\
\objectname[]{MRC~0406$-$244} &0406$-$244& 19.11 & 2.440 & 2000 Sep 15 & 6000 & $-$48 & 0.6 \\
\objectname[]{B3~0731$+$438 } &0731$+$438& 19.35 & 2.429 & 2000 Dec 17 & 4000 & $+$18 & 0.6 \\
\objectname[]{TXS~0828$+$193} &0828$+$193& 19.40 & 2.572 & 2000 Dec 18 & 4000 & $+$38 & 0.7 \\
\objectname[]{4C~$-$00.54   } &1410$-$001& 19.48 & 2.363 & 2000 May 20 & 3000 & $-$47 & 0.6 \\
\objectname[]{4C~$-$00.62   } &1558$-$003& 19.69 & 2.527 & 2000 May 21 & 4000 & $+$76 & 0.6 \\
\objectname[]{4C~+10.48     } &1707$+$105& 20.72 & 2.349 & 2001 Aug 01 & 4000 & $+$52 & 0.8 \\
\objectname[]{3C 362        } &1744$+$183& 20.67 & 2.281 & 2000 May 22 & 3000 & $+$51 & 0.7 \\
\objectname[]{4C~+40.36     } &1809$+$407& 18.93 & 2.265 & 2000 May 23 & 3200 & $+$82 & 0.8 \\
\objectname[]{4C~+48.48     } &1931$+$480& 19.86 & 2.343 & 2000 Jul 24 & 4000 & $+$29 & 1.0 \\
\objectname[]{MRC~2104$-$242} &2104$-$242& 21.13 & 2.491 & 2000 Sep 14 & 4000 & $+$17 & 0.8 \\
\objectname[]{4C~+23.56     } &2105$+$233& 20.22 & 2.483 & 2000 Jul 24 & 1000\tablenotemark{b} & $+$51 & 0.8 \\
\enddata
\tablenotetext{a}{Observed $H$-band magnitudes by OHS imaging observations with a 2\farcs2-diameter aperture. Typical photometric errors are 0.05 mag.}
\tablenotetext{b}{The exposure sequence was terminated by instrument trouble with the telescope.}
\end{deluxetable}
\clearpage
\begin{deluxetable}{llcccccc}
\tabletypesize{\normalsize}
\tablecaption{\sc Emission Line Properties and Continuum Levels}
\tablewidth{0pt}
\tablehead{\colhead{} & \colhead{} & \colhead{Wavelength} & \colhead{} & \colhead{Line} & \colhead{Equivalent} & \colhead{FWHM\tablenotemark{a,d}} & \colhead{Continuum} \\
\colhead{Common Name} & \colhead{Lines} & \colhead{($\mu$m)} & \colhead{Redshift\tablenotemark{a}} & \colhead{Flux\tablenotemark{b}} & \colhead{Width\tablenotemark{c} (\AA)} & \colhead{(km s$^{-1}$)} & \colhead{Level\tablenotemark{e}}}
\startdata
MRC~0156$-$252&[\ion{O}{3}]~$\lambda$5007  & 1.5139 & 2.0237 & 13.02$\pm$0.52 &  109 & 1213$^{+82}_{-36}$& 39.57\\
              &[\ion{Ne}{3}]~$\lambda$3869 & 1.1699 &   ...  &  0.84$\pm$0.52 &   20 &  ...  & 13.94\\
              &[\ion{O}{2}]~$\lambda$3727  & 1.1276 & 2.0256 &  1.55$\pm$0.76 &   44 &  784$^{+917}_{-329}$& 11.75\\
\hline
MRC~0200$+$015&[\ion{O}{3}]~$\lambda$5007  & 1.6184 & 2.2322 & 21.60$\pm$0.34 &  630 & 1628$^{+32}_{-25}$& 10.60\\
              &H$\beta$                   & 1.5722 & 2.2342 &  1.97$\pm$0.34 &   51 & 1402$^{+303}_{-234}$& 12.00\\
              &\ion{He}{2}~$\lambda$4686   & 1.5157 & 2.2346 &  0.78$\pm$0.52 &   17 & 1492$^{+6938}_{-561}$& 13.86\\
              &[\ion{Ne}{3}]~$\lambda$3869 & 1.2522 & 2.2365 &  2.60$\pm$0.42 &   97 & 2189$^{+468}_{-368}$&  8.27\\
              &[\ion{O}{2}]~$\lambda$3727  & 1.2033 & 2.2286 &  6.92$\pm$0.52 &  282 & 2045$^{+188}_{-167}$&  7.59\\
\hline
TXS~0211$-$122&[\ion{O}{3}]~$\lambda$5007  & 1.6727 & 2.3407 & 12.18$\pm$0.31 &  545 & 1164$^{+21}_{-59}$&  6.69\\
              &H$\beta$                   & 1.6220 & 2.3367 &  0.83$\pm$0.40 &   35 &  889$^{+681}_{-316}$&  7.10\\
              &\ion{He}{2}~$\lambda$4686   & 1.5655 &   ...  &  0.10$\pm$0.36 &    4 &  ...  &  7.66\\
              &[\ion{Ne}{3}]~$\lambda$3869 & 1.2922 & 2.3399 &  0.91$\pm$0.43 &   42 & 1092$^{+952}_{-343}$&  6.40\\
              &[\ion{O}{2}]~$\lambda$3727  & 1.2449 & 2.3402 &  2.57$\pm$0.42 &  114 & 1622$^{+337}_{-253}$&  6.74\\
\hline
MRC~0214$+$183&[\ion{O}{3}]~$\lambda$5007  & 1.5671 & 2.1299 & 32.69$\pm$0.32 & 1346 & 1673$^{+20}_{-16}$&  7.76\\
              &H$\beta$                   & 1.5221 & 2.1313 &  4.03$\pm$0.44 &  187 & 2272$^{+307}_{-248}$&  6.88\\
              &[\ion{Ne}{3}]~$\lambda$3869 & 1.2109 & 2.1299 &  2.44$\pm$0.51 &   91 & 2571$^{+774}_{-393}$&  8.55\\
              &[\ion{O}{2}]~$\lambda$3727  & 1.1663 & 2.1293 &  5.99$\pm$0.71 &  237 & 2019$^{+292}_{-238}$&  8.06\\
\hline
MRC~0406$-$244&[\ion{O}{3}]~$\lambda$5007  & 1.7161 & 2.4274 & 37.79$\pm$0.34 & 1677 & 1340$^{+44}_{-5}$&  6.57\\
              &H$\beta$                   & 1.6658 & 2.4269 &  3.00$\pm$0.21 &  121 & 1389$^{+116}_{-103}$&  7.23\\
              &\ion{He}{2}~$\lambda$4686   & 1.6046 & 2.4243 &  0.63$\pm$0.18 &   24 & 1208$^{+829}_{-360}$&  7.60\\
              &[\ion{Ne}{3}]~$\lambda$3869 & 1.3252 & 2.4251 &  2.89$\pm$0.39 &   95 & 1659$^{+244}_{-207}$&  8.93\\
              &[\ion{O}{2}]~$\lambda$3727  & 1.2778 & 2.4285 &  7.69$\pm$0.67 &  252 & 1786$^{+166}_{-149}$&  8.91\\
              &[\ion{Ne}{5}]~$\lambda$3426 & 1.1723 & 2.4216 &  1.59$\pm$0.29 &   52 & 1661$^{+478}_{-264}$&  9.01\\
\hline
B3~0731$+$438 &[\ion{O}{3}]~$\lambda$5007  & 1.7183 & 2.4319 & 31.36$\pm$0.43 & 1319 & 1135$^{+22}_{-14}$&  6.93\\
              &H$\beta$                   & 1.6690 & 2.4335 &  2.61$\pm$0.34 &  110 & 1179$^{+183}_{-155}$&  6.94\\
              &\ion{He}{2}~$\lambda$4686   & 1.6097 & 2.4352 &  0.37$\pm$0.33 &   17 & 1907$^{+7437}_{-1022}$&  6.25\\
              &[\ion{Ne}{3}]~$\lambda$3869 & 1.3269 & 2.4296 &  2.15$\pm$0.43 &  125 & 2059$^{+585}_{-418}$&  5.01\\
              &[\ion{O}{2}]~$\lambda$3727  & 1.2792 & 2.4324 &  5.54$\pm$0.69 &  290 & 1508$^{+188}_{-161}$&  5.56\\
              &[\ion{Ne}{5}]~$\lambda$3426 & 1.1751 & 2.4300 &  0.82$\pm$0.52 &   36 & 2334$^{+3431}_{-959}$&  6.72\\
\hline
TXS~0828$+$193&[\ion{O}{3}]~$\lambda$5007  & 1.7888 & 2.5726 & 35.64$\pm$0.36 & 1571 & 1434$^{+53}_{-6}$&  6.35\\
              &H$\beta$                   & 1.7388 & 2.5770 &  3.74$\pm$0.24 &  165 & 1448$^{+112}_{-100}$&  6.32\\
              &\ion{He}{2}~$\lambda$4686   & 1.6773 & 2.5793 &  1.10$\pm$0.20 &   53 & 2060$^{+495}_{-363}$&  5.84\\
              &H$\gamma$+[\ion{O}{3}]     & 1.5553 & 2.5746\tablenotemark{f} &  2.35$\pm$0.24 &  105 & 2465$^{+314}_{-262}$&  6.24\\
              &[\ion{O}{2}]~$\lambda$3727  & 1.3314 & 2.5723 &  6.55$\pm$0.23 &  223 & 2134$^{+86}_{-81}$&  8.22\\
              &[\ion{Ne}{5}]~$\lambda$3426 & 1.2257 & 2.5776 &  2.85$\pm$0.30 &   76 & 2302$^{+245}_{-318}$& 10.43\\
\hline
4C~$-$00.54   &[\ion{O}{3}]~$\lambda$5007  & 1.6852 & 2.3657 & 28.28$\pm$0.96 & 1251 & 1298$^{+53}_{-48}$&  6.72\\
              &H$\beta$                   & 1.6356 & 2.3647 &  1.15$\pm$1.16 &   42 & 1021$^{+9188}_{-494}$&  8.10\\
              &\ion{He}{2}~$\lambda$4686   & 1.5772 &   ...  &  0.21$\pm$1.48 &    8 &  ...  &  8.29\\
              &[\ion{Ne}{3}]~$\lambda$3869 & 1.3022 &   ...  &  1.10$\pm$1.34 &   29 &  ...  & 11.28\\
              &[\ion{O}{2}]~$\lambda$3727  & 1.2540 & 2.3648 &  3.79$\pm$1.15 &  123 & 1166$^{+515}_{-311}$&  9.18\\
\hline
4C~$-$00.62   &[\ion{O}{3}]~$\lambda$5007  & 1.7666 & 2.5282 &  6.54$\pm$0.59 &  114 &  856$^{+87}_{-72}$& 16.29\\
              &H$\beta$                   & 1.7153 & 2.5287 &  0.99$\pm$0.36 &   19 & 1185$^{+775}_{-374}$& 14.47\\
              &\ion{He}{2}~$\lambda$4686   & 1.6533 &   ...  &  0.00$\pm$0.50 &    0 &  ...  & 12.67\\
              &[\ion{O}{2}]~$\lambda$3727  & 1.3155 & 2.5295 &  2.07$\pm$0.35 &   65 & 1301$^{+344}_{-240}$&  9.08\\
              &[\ion{Ne}{5}]~$\lambda$3426 & 1.2088 &   ...  &  0.86$\pm$0.41 &   30 &  ...  &  8.13\\
\hline
4C~+10.48     &[\ion{O}{3}]~$\lambda$5007  & 1.6766 & 2.3486 &  5.62$\pm$0.42 &  615 & 1309$^{+142}_{-105}$&  2.73\\
              &H$\beta$                   & 1.6288 & 2.3508 &  1.02$\pm$0.52 &   91 & 1466$^{+1041}_{-550}$&  3.34\\
              &\ion{He}{2}~$\lambda$4686   & 1.5685 & 2.3472 &  0.67$\pm$0.36 &   55 &  647$^{+479}_{-218}$&  3.65\\
              &[\ion{Ne}{3}]~$\lambda$3869 & 1.2956 &   ...  &  0.43$\pm$0.66 &   33 &  ...  &  3.95\\
              &[\ion{O}{2}]~$\lambda$3727  & 1.2476 & 2.3475 &  1.98$\pm$0.64 &  140 & 1589$^{+807}_{-453}$&  4.22\\
\hline
3C 362        &[\ion{O}{3}]~$\lambda$5007  & 1.6442 & 2.2839 &  5.35$\pm$0.28 &  587 & 1397$^{+71}_{-100}$&  2.78\\
              &H$\beta$                   & 1.5954 & 2.2820 &  0.42$\pm$0.25 &   38 &  766$^{+1439}_{-376}$&  3.33\\
              &\ion{He}{2}~$\lambda$4686   & 1.5388 &   ...  &  0.40$\pm$0.44 &   29 &  ...  &  4.11\\
              &[\ion{Ne}{3}]~$\lambda$3869 & 1.2705 &   ...  &  0.02$\pm$0.52 &    2 &  ...  &  4.25\\
              &[\ion{O}{2}]~$\lambda$3727  & 1.2221 & 2.2790 &  1.23$\pm$0.34 &  101 & 1099$^{+368}_{-247}$&  3.71\\
\hline
4C~+40.36     &[\ion{O}{3}]~$\lambda$5007  & 1.6360 & 2.2675 & 35.29$\pm$0.59 &  832 & 1987$^{+28}_{-52}$& 12.98\\
              &H$\beta$                   & 1.5905 & 2.2720 &  3.24$\pm$0.48 &   74 & 1723$^{+289}_{-228}$& 13.33\\
              &\ion{He}{2}~$\lambda$4686   & 1.5328 & 2.2709 &  0.76$\pm$0.61 &   17 & 1823$^{+5541}_{-1126}$& 13.91\\
              &[\ion{Ne}{3}]~$\lambda$3869 & 1.2640 & 2.2670 &  3.10$\pm$0.81 &   89 & 1741$^{+594}_{-378}$& 10.64\\
              &[\ion{O}{2}]~$\lambda$3727  & 1.2179 & 2.2677 & 13.89$\pm$0.73 &  406 & 2390$^{+151}_{-139}$& 10.47\\
\hline
4C~+48.48     &[\ion{O}{3}]~$\lambda$5007  & 1.6749 & 2.3452 & 15.57$\pm$0.21 &  978 & 1243$^{+29}_{-10}$&  4.76\\
              &H$\beta$                   & 1.6246 & 2.3422 &  1.96$\pm$0.28 &  111 & 1795$^{+269}_{-222}$&  5.28\\
              &\ion{He}{2}~$\lambda$4686   & 1.5676 &   ...  &  0.47$\pm$0.20 &   25 &  ...  &  5.60\\
              &[\ion{Ne}{3}]~$\lambda$3869 & 1.2953 & 2.3479 &  0.83$\pm$0.22 &   44 & 1044$^{+425}_{-223}$&  5.58\\
              &[\ion{O}{2}]~$\lambda$3727  & 1.2473 & 2.3467 &  3.98$\pm$0.24 &  232 & 1693$^{+122}_{-111}$&  5.12\\
\hline
MRC~2104$-$242&[\ion{O}{3}]~$\lambda$5007  & 1.7469 & 2.4889 &  5.73$\pm$0.18 & 2486 & 1220$^{+52}_{-45}$&  0.66\\
              &H$\beta$                   & 1.6968 & 2.4907 &  0.88$\pm$0.23 &  307 & 1423$^{+746}_{-423}$&  0.82\\
              &\ion{He}{2}~$\lambda$4686   & 1.6349 &   ...  &  0.23$\pm$0.22 &   76 &  ...  &  0.87\\
              &[\ion{O}{2}]~$\lambda$3727  & 1.3012 & 2.4913 &  0.97$\pm$0.22 & 1459 & 1982$^{+809}_{-514}$&  0.19\\
              &[\ion{Ne}{5}]~$\lambda$3426 & 1.1953 &   ...  &  0.14$\pm$0.25 &  ...\tablenotemark{g} &  ...  &  ...\tablenotemark{g} \\
\hline
4C~+23.56     &[\ion{O}{3}]~$\lambda$5007  & 1.7440 & 2.4832 &  7.33$\pm$0.58 &  474 & 1258$^{+105}_{-122}$&  4.44\\
              &H$\beta$                   & 1.6985 & 2.4941 &  1.88$\pm$0.96 &  102 & 3047$^{+2989}_{-1130}$&  5.27\\
              &\ion{He}{2}~$\lambda$4686   & 1.6322 &   ...  &  0.90$\pm$0.70 &   44 &  ...  &  5.93\\
              &[\ion{O}{2}]~$\lambda$3727  & 1.2982 &   ...  &  1.47$\pm$0.74 &   69 &  ...  &  6.13\\
\enddata
\tablenotetext{a}{For faint emission lines represented by an ellipsis, we assume that the redshift is the same as 
that for [\ion{O}{3}]~$\lambda$5007 and that the FWHM is the instrumental value of $\sim$75~\AA.}
\tablenotetext{b}{The flux with 1~$\sigma$ errors, in units of 10$^{-16}$~erg~s$^{-1}$~cm$^{-2}$.}
\tablenotetext{c}{The rest-frame equivalent width. The signal-to-noise ratios are almost the same as that of the flux.}
\tablenotetext{d}{No correction has been made for the instrumental width of $\sim$75~\AA, which corresponds to
$\sim$1400~km s$^{-1}$ at 1.60~$\mu$m and $\sim$1800~km s$^{-1}$ at 1.25~$\mu$m. Note that the typical seeing size ($\sim 0\farcs 7$)
is smaller than the slit width (1\arcsec), causing smaller line width than the instrumental value.}
\tablenotetext{e}{The flux density of the continuum at the line position in units of 10$^{-15}$~erg~s$^{-1}$~cm$^{-2}$~$\mu$m$^{-1}$.}
\tablenotetext{f}{The average redshift for H$\gamma$ and [\ion{O}{3}]~$\lambda$4363.}
\tablenotetext{g}{The continuum level becomes a negative value.}
\end{deluxetable}
\clearpage
\begin{deluxetable}{lcccl}
\tabletypesize{\normalsize}
\tablecaption{\sc UV Line Flux}
\tablewidth{0pt}
\tablehead{\colhead{Common Name} & \colhead{\ion{C}{4}~$\lambda$1549\tablenotemark{a}} & \colhead{\ion{He}{2}~$\lambda$1640\tablenotemark{a}} & \colhead{\ion{C}{3}]~$\lambda$1909\tablenotemark{a}} & \colhead{Reference}}
\startdata
\objectname[]{MRC~0200$+$015} &   4.2$\pm$0.5  &   3.2$\pm$0.4  &   4.0$\pm$0.5 &1\\
\objectname[]{TXS~0211$-$122} &  2.82$\pm$0.10 &  1.57$\pm$0.03 &  0.74$\pm$0.05&2\\
\objectname[]{MRC~0214$+$183} &   3.0$\pm$0.3  &   1.8$\pm$0.2  &   1.8$\pm$0.2 &1\\
\objectname[]{MRC~0406$-$244} &   4.0$\pm$0.3  &   5.4$\pm$0.3  &   4.9$\pm$0.5 &3,4\\
\objectname[]{B3~0731$+$438 } &  4.65$\pm$0.10 &  3.04$\pm$0.05 &  2.12$\pm$0.05&2\\
\objectname[]{TXS~0828$+$193} & 18.17$\pm$0.45 & 10.38$\pm$0.14 &  5.59$\pm$0.68&2\\
\objectname[]{4C~$-$00.54   } &  2.91$\pm$0.20 &  2.15$\pm$0.05 &  0.96$\pm$0.07&2\\
\objectname[]{4C~$-$00.62   } &   2.7$\pm$0.3  &   1.7$\pm$0.2  &   1.2$\pm$0.2 &1\\
\objectname[]{4C~+10.48     } &   0.2$\pm$0.2  &   0.9$\pm$0.2  &   0.2$\pm$0.1 &1\\
\objectname[]{4C~+40.36     } & 13.61$\pm$0.51 &  7.77$\pm$0.56 &  7.39$\pm$1.18&2\\
\objectname[]{4C~+48.48     } &  5.55$\pm$0.05 &  5.32$\pm$0.35 &  3.28$\pm$0.10&2\\
\objectname[]{MRC~2104$-$242} &   2.4$\pm$0.3  &   3.0$\pm$0.3  &   2.1$\pm$0.9 &5,6\\
\objectname[]{4C~+23.56     } &  1.80$\pm$0.15 &  1.50$\pm$0.10 &  1.06$\pm$0.10&2\\
\enddata
\tablenotetext{a}{The flux with 1$\sigma$ errors, in units of 10$^{-16}$~erg~s$^{-1}$~cm$^{-2}$.}
\tablerefs{(1) R{\"o}ttgering et al. 1997; (2) Vernet et al. 2001; (3) Rush et al. 1997; (4) Taniguchi et al. 2001; (5) Villar-Mart{\'i}n et al. 1999; (6) Overzier et al. 2001.}
\tablecomments{The UV line flux taken from the past literature. The typical slit width is $1\arcsec- 2\arcsec$, and the spatial extent of the aperture is large enough to cover all the flux of emission lines.}
\end{deluxetable}
\clearpage
\begin{deluxetable}{lcccccccc}
\tabletypesize{\normalsize}
\tablecaption{\sc Best Fit Parameters}
\tablewidth{0pt}
\tablehead{\colhead{} & \multicolumn{5}{c}{Single-Photoionization Model} & \multicolumn{3}{c}{Shock+Precursor model}\\
\colhead{Common Name} & \colhead{$\chi_{\rm ph}^2$\tablenotemark{a}} & \colhead{$\alpha$} & \colhead{$Z$} & \colhead{log$U$} & \colhead{log$N_{\rm H}$\tablenotemark{b}} & \colhead{$\chi_{\rm s+p}^2$\tablenotemark{a}} & \colhead{$B/\sqrt{n}$} & \colhead{$v$}}
\startdata
\objectname[]{MRC~0200$+$015} &  2.94 & $-$0.615 & 0.260 & $-$2.415 & 2.00 & 41.8 & ...   & ...\\
\objectname[]{TXS~0211$-$122} &  30.3 & ...      & ...   & ...      & ...  & 6.45 & 0.245 & 500\tablenotemark{c}\\
\objectname[]{MRC~0214$+$183} &  2.94\tablenotemark{d} & $-$0.500 & 0.183 & $-$2.269 & 3.47 & 34.6\tablenotemark{d} & ...   & ...\\
\objectname[]{MRC~0406$-$244} &  14.2 & $-$1.176 & 0.530 & $-$2.323 & 1.00\tablenotemark{c} & 115  & ...   & ...\\
\objectname[]{B3~0731$+$438 } &  41.0\tablenotemark{e}& ... & ... & ... & ... & 154  & ...   & ...\\
\objectname[]{TXS~0828$+$193} &  117\tablenotemark{e} & ... & ... & ... & ... & 251  & ...   & ...\\
\objectname[]{4C~$-$00.54   } &  2.40 & $-$0.896 & 3.000\tablenotemark{c} & $-$1.699 & 1.00\tablenotemark{c} & 1.42 & 0.013 & 481\\
\objectname[]{4C~$-$00.62   } &  6.18 & $-$0.172 & 0.131 & $-$2.070 & 3.00 & 7.15 & 0.637 & 500\tablenotemark{c}\\
\objectname[]{4C~+10.48     } &  4.41 & $-$1.250 & 1.762 & $-$2.477 & 2.00 & 21.2 & ...   & ...\\
\objectname[]{4C~+40.36     } &  17.4 & $-$0.402 & 0.278 & $-$2.272 & 2.00 & 19.7 & ...   & ...\\
\objectname[]{4C~+48.48     } &  57.8\tablenotemark{e} & ... & ... & ... & ... & 162  & ...   & ...\\
\objectname[]{MRC~2104$-$242} &  2.94 & $-$0.660 & 0.147 & $-$2.554 & 4.00\tablenotemark{c} & 6.64 & 0.000\tablenotemark{c} & 440\\
\objectname[]{4C~+23.56     } &  0.06 & $-$0.566 & 0.175 & $-$2.283 & 3.21 & 26.0 & ...   & ...\\
\enddata
\tablenotetext{a}{The $\chi^2$ value for the best fit parameters of each model, photoionization ($\chi_{\rm ph}^2$) or shock+precursor ($\chi_{\rm s+p}^2$), respectively.}
\tablenotetext{b}{The hydrogen density, which is not as sensitive to the emission line ratio compared to the other parameters (see $\S3.2.1$).}
\tablenotetext{c}{Maximum or minimum value of the domain of the parameter.}
\tablenotetext{d}{Degree of freedom is smaller than the others because \ion{He}{2}~$\lambda$4686 was out of the observed range for this object.}
\tablenotetext{e}{The best fit models were rejected for these results.}
\end{deluxetable}
\clearpage
\begin{deluxetable}{ccccc}
\tabletypesize{\normalsize}
\tablecaption{\sc Significance Level to reject Zero Correlation Hypothesis}
\tablewidth{0pt}
\tablehead{\colhead{Correlating Value} & \colhead{log$P_{opt}$\tablenotemark{a}} & \colhead{$\alpha_{opt}$\tablenotemark{b}} & \colhead{log$L_{[OIII]}$} & \colhead{logFWHM\tablenotemark{c}}}
\startdata
log$P_{radio}$\tablenotemark{d}&{\rm 0.797(+0.08)}&{\rm 0.355($-$0.28)}&{\rm 0.838($-$0.06)}&{\rm 0.781($-$0.08)}\\
log$D_{radio}$\tablenotemark{e}&{\rm 0.164($-$0.39)}&{\bf 0.040(+0.57)}&{\rm 0.214($-$0.35)}&0.089($-$0.47)\\
$\alpha_{radio}$&{\rm 0.385($-$0.25)}&{\rm 0.521(+0.20)}&{\rm 0.498($-$0.20)}&{\rm 0.484($-$0.20)}\\
log$P_{opt}$&  ... &{\bf 0.003($-$0.75)}&0.055(+0.52)&{\rm 0.871($-$0.05)}\\
$\alpha_{opt}$&  ... &  ... &{\rm 0.760($-$0.09)}&{\rm 0.558(+0.18)}\\
log$L_{[OIII]}$&  ... &  ... &  ... &{\rm 0.173(+0.39)}\\
\enddata
\tablenotetext{a}{Continuum brightness at the position of [\ion{O}{3}]~$\lambda$5007 emission line.}
\tablenotetext{b}{Continuum slope between [\ion{O}{2}]~$\lambda$3727 and [\ion{O}{3}]~$\lambda$5007 (see Figure~1). MRC~2104$-$242 is not included in this column because the observed $J$-band continuum is too faint to determine the reliable color.}
\tablenotetext{c}{FWHM of [\ion{O}{3}] emission line without correction of the instrumental width.}
\tablenotetext{d}{Radio power at rest-178~MHz, interpolated from the observed-408~MHz and 1.4~GHz fluxes.}
\tablenotetext{e}{Radio size \citep{car97,pen99,pen00}.}
\tablecomments{The significance level at which the null hypothesis of zero correlation is disproved. Smaller value indicates a significant correlation (bold values). Values in parentheses are the correlation coefficients. MRC~0156$-$252 is not included in these calculations because this object is expected to be a dusty quasar, showing different characteristics from the other sample (see Appendix A.1.).}
\end{deluxetable}
\end{document}